\documentclass[12pt]{article}
\usepackage{amsmath, amsthm}
\usepackage{setspace}
\usepackage{amsfonts}
\numberwithin{equation}{section}
\usepackage{amsfonts}
 \usepackage[hang,small,bf]{caption}
\usepackage{subfigure}
\usepackage[bottom]{footmisc}
\usepackage{fullpage}

\title{
\textmd{Transition Amplitudes in de Sitter Space}
\vspace{5 mm}}
\author{Matthew Dodelson\footnote{mdodelso@stanford.edu}\vspace{5 mm}\\\emph{  Department of Physics, Brown University}\\
\emph{Providence, RI 02912, USA}}
\date{
}
\begin{document}
\maketitle\vspace{20 mm}
\abstract{Maldacena has shown that the wavefunction of the universe in de Sitter space can be viewed as the partition function of a conformal field theory. In this paper, we investigate this approach to the dS/CFT correspondence in further detail. We emphasize that massive bulk fields are dual to two primary operators on the boundary, which encode information about the two independent behaviors of bulk expectation values at late times. An operator statement of the duality is given, and it is shown that the resulting boundary correlators can be interpreted as transition amplitudes from the Bunch-Davies vacuum to an excited state in the infinite future. We also explain how these scattering amplitudes can be used to compute late-time Bunch-Davies expectation values, and comment on the effects of anomalies in the dual CFT on such expectation values.}\thispagestyle{empty}
\pagebreak

\tableofcontents

\setcounter{page}{2}

\section{Introduction}

In \cite{maldacena}, it was proposed that the Hartle-Hawking wavefunction \cite{hartle} in de Sitter space is equal to the partition function of a conformal field theory on the boundary. If such a dual field theory exists, then it gives a powerful method for computing in-in expectation values of bulk fields in the Euclidean vacuum at late times:  instead of doing a path integral over the standard Schwinger-Keldysh contour, one can directly evaluate the wavefunction by computing the CFT correlators. Expectation values can then be calculated by squaring the wavefunction, inserting a product of bulk fields, and integrating over the CFT sources. \\
\indent This procedure was investigated in \cite{maldacena} in the context of a model of inflation with only massless fields. The massless scalar field operator in de Sitter space has only one independent behavior near future null infinity $\mathcal{I}^+$ that can be identified with a primary operator in a CFT, implying that a massless scalar in the bulk corresponds to only one primary operator on the boundary\footnote{In fact, one can identify both independent behaviors of a massless bulk scalar field with CFT operators, but one of the operators has logarithmic correlators and is therefore not primary \cite{turok}. }. In this paper, we will generalize the results of \cite{maldacena} to the case where the bulk field is massive, and has two leading behaviors that can be identified with primary operators. This leads to a CFT correlator structure that is richer than in the massless case, since bulk correlation functions have more than one independent behavior near $\mathcal{I}^+$.\\
\indent Several important aspects of the duality were recently clarified in \cite{harlow}. In particular, it was shown that extrapolating bulk correlation functions to the boundary does not give the CFT correlation functions defined in \cite{maldacena}. One of the goals of this work is to represent the correlators of \cite{maldacena} as extrapolated transition amplitudes between the Bunch-Davies vacuum and a certain final state, with a time-ordered product of fields inserted in between. This allows for an operator formulation of the correspondence, where the boundary operators are defined by pulling the bulk operators to the boundary, as in AdS/CFT \cite{balasubramanian}.\\
\indent The rest of this paper is organized as follows. In Section \ref{seccorrelation}, we define the CFT operators in terms of the boundary behavior of the bulk fields, and express the correlators of \cite{maldacena} as the boundary values of transition amplitudes. We also prove that modifying the final state in the amplitudes is accomplished by deforming the CFT by a certain marginal operator. In Section \ref{secscattering}, we reinterpret the correlation functions as a scattering matrix, and derive a simple diagrammatic expansion for this S-matrix. We make contact with the in-in formalism in Section \ref{secinin}, and show that Bunch-Davies expectation values at late times are computed by integrating over two sources in the dual CFT. Section \ref{secanomalies} then concludes this work by discussing the effects of the conformal anomaly in the dual field theory on bulk expectation values. We argue that such anomalies are only visible in expectation values involving time derivatives of the bulk field $\phi$.
\section{Correlation Functions}
\label{seccorrelation}
The goal of this section is to express the correlation functions of \cite{maldacena} in terms of bulk transition amplitudes. For simplicity, we will consider a scalar field $\phi$ in the bulk, governed by the action
\begin{align}
S=-\frac{1}{2}\int d^{d+1}x\, \sqrt{-g}\left((\nabla \phi)^2+m^2\phi^2\right)+S_{\text{int}}.
\end{align}
We work in the inflationary patch of $dS_{d+1}$, where the metric reads
\begin{align}
ds^2=\frac{-d\eta^2+d\vec{x}^2}{\eta^2},\hspace{10 mm}-\infty<\eta<0.
\end{align}
Solutions to the equations of motion behave near $\mathcal{I}^+$ as \cite{dscft}
\begin{align}
\phi(\eta,\vec{x})\sim \phi_+(\vec{x})(-\eta)^{\Delta_-}+\phi_-(\vec{x})(-\eta)^{\Delta_+},\,
\end{align}
where the dimensions are
\begin{align}
\Delta_{\pm}=\frac{d}{2}\pm \nu,\hspace{10 mm}\nu=\frac{1}{2}\sqrt{d^2-4m^2}.
\end{align}
We restrict our analysis to the case where $\Delta_{\pm}$ are real, which is equivalent to only considering sufficiently light scalars. \\
\indent In anti-de Sitter space, the free field operator is expressed as a sum over normalizable modes, which are fast-falling at the boundary (see \cite{balasubramanian3}, for example). On the other hand, the field operator in de Sitter space has two behaviors at $\mathcal{I}^+$, and each of these can be identified with a CFT operator on the boundary. To be precise, we postulate the operator relation
\begin{align}
\Phi(\eta,\vec{x})\sim -\frac{i}{2\nu}\mathcal{O}_-(\vec{x})(-\eta)^{\Delta_-}+\frac{i}{2\nu}\mathcal{O}_+(\vec{x})(-\eta)^{\Delta_+}.\label{postulate}
\end{align}
Here we have chosen a convenient normalization for $\mathcal{O}_\pm $, which agrees with the AdS/CFT normalization up to a factor of $\pm i$ \cite{balasubramanian,polchinskireview}. Note that $\mathcal{O}_+$ and $\mathcal{O}_-$  are anti-Hermitian. \\
\indent It is important to emphasize that we are not assuming the existence of a dual CFT. Rather, the claim is simply that expectation values of $\mathcal{O}_{\pm}$ between bulk vacuum states obey the same constraints as correlation functions in a CFT. Also, note that our postulate (2.5) differs in spirit from much of the recent literature, which advocates a dS/CFT dictionary that is the analytic continuation from AdS/CFT \cite{maldacena, harlow,strominger}. One could imagine fixing future boundary conditions on a bulk field in de Sitter space by analytically continuing the field operator from AdS, but we instead find it convenient to fix boundary conditions by fixing the final state. The relation between the two approaches will be clarified below; it turns out that the correlation functions of \cite{maldacena,harlow,strominger} are equal to the expectation values of $\mathcal{O}_+$ between two certain bulk vacua. The operator $\mathcal{O}_-$ acts as the source in the CFT of \cite{maldacena,harlow,strominger}, and captures the leading behaviors of expectation values.\\
\indent By dimensional analysis, we expect that $\mathcal{O}_{\pm}$ has conformal dimension $\Delta_{\pm}$. To confirm this, let us expand the operator $\Phi$ in terms of Bunch-Davies creation and annihilation operators \cite{birrell},
\begin{align}
\Phi(\eta,\vec{k})=\frac{\sqrt{\pi}(-\eta)^{d/2}}{2}\left(H_\nu(\eta k)^*a(\vec{k})+H_\nu(\eta k)a^\dagger(-\vec{k})\right),\label{phiex}
\end{align}
where $H_\nu$ is a Hankel function of the first kind. Taking $\eta\to 0$, we find
\begin{align}
&\mathcal{O}_+(\vec{k})=\frac{k^\nu\Gamma(1-\nu)}{2^{\nu} \sqrt{\pi}}(a^\dagger(-\vec{k})-a(\vec{k})),\label{op}\\
&\mathcal{O}_-(\vec{k})=\frac{k^{-\nu}\Gamma(1+\nu)}{2^{-\nu}\sqrt{\pi}}(e^{-i\nu \pi}a^\dagger(-\vec{k})-e^{i\nu \pi}a(\vec{k})),\label{om}
\end{align}
and these indeed transform as operators of dimension $\Delta_{\pm}$. In deriving the expressions (\ref{op}) and (\ref{om}), we have assumed that $\nu$ is neither an integer nor a half-integer -- these cases, which include the case of a massless field, must be treated separately.\\
\indent Now let us define a state $|+\rangle$ that is annihilated by the slow-falling modes near $\mathcal{I}^+$, so that $\mathcal{O}_-|+\rangle=0$ \cite{strominger2}. Suppose that we are interested in computing transition amplitudes from the Bunch-Davies vacuum $|\text{BD}\rangle$ to $\langle +|$, with a time-ordered product of $\Phi$ fields inserted. These quantities can be computed via a path integral over fields obeying fixed boundary conditions in the infinite past and future,
\begin{align}
\langle +|\text{T}\Phi(x_1)\cdots \Phi(x_n)|\text{BD}\rangle=\int_{\phi(\eta=-\infty)\sim e^{ik\eta}}^{\phi(\eta=0)\sim(-\eta)^{\Delta_+}}[d\phi]\, \phi(x_1)\cdots \phi(x_n)\exp(iS[\phi]),\label{transition}
\end{align}
where we have absorbed an overall normalization into the path integral measure. The boundary conditions at early times require that $\phi$ is purely negative frequency with respect to $|\text{BD}\rangle$ \cite{maldacena}, while the late-time boundary conditions ensure that $\phi$ is fast-falling near the boundary. \\
\indent In \cite{maldacena,harlow}, it was shown that the infrared wavefunction in AdS analytically continues to the Bunch-Davies wavefunction in de Sitter space. Similarly, we see that the matrix elements (\ref{transition}) are the analytic continuations of bulk correlation functions from AdS  (related claims have appeared in \cite{harlow,strominger3}). Indeed, the boundary conditions in (\ref{transition}) analytically continue to the AdS boundary conditions under $\eta=iz$. Another way to see this is by computing the Green's function $\langle +|\text{T}\Phi(x)\Phi(y)|\text{BD}\rangle$, since the amplitudes (\ref{transition}) are evaluated in perturbation theory using this Green's function. Using (\ref{phiex}) and the fact that $\mathcal{O}_-|+\rangle=0$, we find
\begin{align}
\langle +|\text{T}\Phi(\eta,\vec{k})\Phi(\eta',\vec{k}')|\text{BD}\rangle&=-\frac{\pi  }{2}(\eta\eta')^{d/2}\left[J_\nu(\eta k)H_\nu(\eta' k)\theta(\eta-\eta')\right.\notag\\
&\hspace{15 mm}\left.+H_\nu(\eta k)J_\nu(\eta' k)\theta(\eta'-\eta)\right](2\pi)^d\delta^d(\vec{k}+\vec{k}'),
\end{align}
which is indeed the analytic continuation of the bulk-to-bulk propagator from AdS \cite{muck}. This explains why the AdS propagator does not analytically continue to the two-point function in any de Sitter invariant vacuum, as was shown indirectly in \cite{bousso,spradlin}. The quantities (\ref{transition}) were referred to in \cite{strominger3} as correlators with future boundary conditions. \\
\indent By taking the arguments of the amplitudes (\ref{transition}) to the boundary and using the operator relation (\ref{postulate}), we may compute the matrix elements of $\mathcal{O}_{\pm}$ between $\langle +|$ and $|\text{BD}\rangle$. In fact, we claim that the matrix elements of $\mathcal{O}_+$ are precisely the CFT correlators defined in \cite{maldacena}. To see this, let us define the state 
\begin{align}
|\phi_+\rangle=\exp\left(-\int\frac{d^d\vec{k}}{(2\pi)^d}\phi_+(\vec{k})\mathcal{O}_+(-\vec{k})\right)|+\rangle,
\end{align}
where $\phi_+(\vec{x})$ is an arbitrary real function. Using the commutation relation
\begin{align}
[\mathcal{O}_+(\vec{k}),\mathcal{O}_-(\vec{k}')]=2i\nu (2\pi)^d\delta^d(\vec{k}+\vec{k}'),
\end{align}
one may check that $|\phi_+\rangle$ is an eigenstate for $\mathcal{O}_-$,
\begin{align}
\mathcal{O}_-(\vec{k})|\phi_+\rangle= 2i \nu\phi_+(\vec{k})|\phi_+\rangle.
\end{align}
The generating functional for matrix elements of $\mathcal{O}_{\pm}$ between $\langle +|$ and $|\text{BD}\rangle$ is therefore
\begin{align}
Z_{\text{CFT}}[\phi_{\pm}]&=\langle +|\exp\left(\int\frac{d^d\vec{k}}{(2\pi)^d}\phi_+(\vec{k})\mathcal{O}_+(-\vec{k})\right)\exp\left(\int\frac{d^d\vec{k}}{(2\pi)^d}\phi_-(\vec{k})\mathcal{O}_-(-\vec{k})\right)|\text{BD}\rangle\notag\\
&=\exp\left(2i\nu\int\frac{d^d\vec{k}}{(2\pi)^d}\phi_+(\vec{k})\phi_-(-\vec{k})\right)\Psi_{\text{BD}}[(-\eta_{\text{c}})^{\Delta_-}\phi_+],\label{generating}
\end{align}
where we used the fact that $\Phi(\eta,\vec{x})\sim -\frac{i}{2\nu}(-\eta)^{\Delta_-}\mathcal{O}_-(\vec{x})$ near $\mathcal{I}^+$. Here we have introduced a late-time cutoff at $\eta=\eta_{\text{c}}$. When the source $\phi_-$ is turned off, we find that the argument of the Bunch-Davies wavefunction acts as a source for an operator of dimension $\Delta_+$, as shown in \cite{maldacena}.\\
\indent Let us now comment on the effect of changing the final state $\langle +|$ to a vacuum state that is instead annihilated by a linear combination of $\mathcal{O}_+$ and $\mathcal{O}_-$. Up to an overall normalization, these vacua can be expressed as the squeezed states \cite{bousso}
\begin{align}
|c\rangle=\exp\left(c\int \frac{d^d\vec{k}}{(2\pi)^d}a^\dagger(\vec{k})a^\dagger(-\vec{k})\right)|+\rangle,
\end{align}  
where $c$ is a constant. Using (\ref{op}) and (\ref{om}), we see that taking $|c\rangle$ instead of $|+\rangle$ as the final state is equivalent to inserting the exponential of a linear combination of the terms
\begin{align}
\int d^d\vec{x}\, \mathcal{O}_+(\vec{x})\mathcal{O}_-(\vec{x}),\hspace{10 mm}\int d^d\vec{x}\, d^d\vec{y}\, \frac{\mathcal{O}_\pm (\vec{x})\mathcal{O}_\pm (\vec{y})}{|\vec{x}-\vec{y}|^{2\Delta_\mp}}\label{operators}
\end{align}
into the generating functional (\ref{generating}). In fact, the operators that are integrated over $\vec{x}$ in (\ref{operators}) are all marginal, so the resulting conformal field theory is a marginal deformation of the original CFT, as conjectured in \cite{bousso}.\footnote{Actually, \cite{bousso} suggests that the marginal operator in question is $\mathcal{O}_+\mathcal{O}_-$, rather than $(a^\dagger)^2$.} Note, however, that the deformation is nonlocal, since the second term in (\ref{operators}) involves products of local operators at different points on the boundary.\\
\indent We glossed over this point above, but it is important to distinguish between the vacua of the interacting and free theories in the expression (\ref{transition}). The Bunch-Davies vacuum that appears as the initial state in (\ref{transition}) is the interacting vacuum, and is related to the vacuum of the free theory by
\begin{align}
|\text{BD}\rangle\propto \text{T}\exp\left(-i\int_{-\infty+i\epsilon}^{0}d\eta\,H_{\text{int}}\right)|\text{BD}\rangle_{\text{free}},
\end{align}
up to an overall factor that was absorbed into the measure in (\ref{transition}). On the other hand, the final state $\langle +|$ is defined the same way in the interacting theory as in the free theory; it is annihilated by $e^{2i\nu\pi}a(\vec{k})-a^\dagger(-\vec{k})$, where $a^\dagger$ and $a$ are the Bunch-Davies creation and annihilation operators that appear in the interaction picture fields.
\section{Scattering Amplitudes}
\label{secscattering}
\indent There is an equivalent way of viewing the correlation functions of $\mathcal{O}_\pm $ in terms of a boundary scattering matrix, along the lines of \cite{spradlin,giddings}. To see this, let us define a pair of solutions to the wave equation by the formulas \nopagebreak 
\begin{align} 
\phi^{\vec{x}}_{\pm}(\eta,\vec{k})=K_\pm (\eta,k)e^{-i\vec{k}\cdot \vec{x}},
\end{align}
where the bulk-to-boundary propagators are given by
\begin{align}
K_\pm (\eta,k)&=\langle \pm |\mathcal{O}_\pm(\vec{k})\Phi(\eta,-\vec{k})|\text{0}\rangle \notag\\
&=\frac{2^{\mp \nu}\pi \eta^{d/2}k^{\pm \nu}}{ i^{d\mp 2\nu-1}\Gamma(\pm \nu)}H_{\pm \nu}(\eta k).
\end{align}
Here $|-\rangle$ is the state that is annihilated by $\mathcal{O}_{+}$. The solutions $\phi^{\vec{x}}_{+}$ and $\phi^{\vec{x}}_{-}$ start out as plane waves that behave as $e^{ik\eta}$, and are fast-falling and slow-falling respectively at $\mathcal{I}^+$, up to a delta function singularity at the point $\vec{x}$,
\begin{align}
\phi_{\pm}^{\vec{x}}(\eta,\vec{k})\sim \left((-\eta)^{\Delta_{\mp}}+\frac{(ik)^{\pm 2\nu}\Gamma(\mp \nu)}{4^{\pm\nu}\Gamma(\pm \nu)}(-\eta)^{\Delta_{\pm}}\right)e^{-i\vec{k}\cdot \vec{x}}.
\end{align}
Note that our definitions of $\phi^{\vec{x}}_{+}$ and $\phi^{\vec{x}}_{-}$ are different from those in \cite{spradlin}; in that work, the bulk-to-boundary propagators were not purely negative frequency at early times. \\
\indent Next, as in \cite{strominger2}, we decompose the field operator $\Phi$ as $\Phi=\Phi_++\Phi_-$, where $\Phi_{\pm}(\eta,\vec{x})\sim (-\eta)^{\Delta_\pm}$ near $\mathcal{I}^+$. It is then straightforward to check that $\mathcal{O}_{\pm}$ can be represented as a creation operator,
\begin{align}
\mathcal{O}_{\pm}(\vec{x})&=(a_{\pm}^{\vec{x}})^{\dagger}\equiv \frac{1}{i}\int_{\eta=\eta_{\text{c}}} {d\Sigma^\mu}\,\phi_{\pm}^{\vec{x}}\overleftrightarrow{\partial_\mu}\Phi_{\pm}.
\end{align}
Here we are assuming that $\Phi$ asymptotes to a free field at $\mathcal{I}^+$; for simplicity, we have left out a wavefunction renormalization for $\Phi$. Thus the matrix elements of $\mathcal{O}_{\pm}$ between $|\text{BD}\rangle$ and an arbitrary final state $\langle f|$ are equal to the scattering amplitudes
\begin{align}
\langle f|\prod_i \mathcal{O}_+(\vec{x}_i)\prod_j \mathcal{O}_-(\vec{y}_j)|\text{BD}\rangle&=\langle f|\prod_i (a_+^{\vec{x}_i})^\dagger \prod_{j}(a^{\vec{y_j}}_-)^\dagger|\text{BD}\rangle.\label{amplitude}
\end{align}
\indent For a generic final state $\langle f|$, one can not directly apply an LSZ-like argument to (\ref{amplitude}), although this can be done by modifying the definition of the bulk-to-boundary propagators. However, in the special case when the expectation values are of the form $\langle \pm|\prod_{i} \mathcal{O}_{\pm}(\vec{x_i})|\text{BD}\rangle$, we may generalize the standard flat-space techniques in order to derive a diagrammatic expansion for the amplitudes, which is analogous to the Witten diagram expansion in AdS \cite{wittenads}. To do this, let us first rewrite the amplitude as
\begin{align}
\langle \pm|\prod_{i} \mathcal{O}_{\pm}(\vec{x_i})|\text{BD}\rangle=\langle \pm|\prod_{j} \frac{1}{i}\int_{\eta=\eta_{\text{c}}} {d\Sigma^\mu}\,\phi_{\pm}^{\vec{x}_j}\overleftrightarrow{\partial_\mu}\Phi|\text{BD}\rangle,
\end{align}
where we have discarded terms that arise from the commutator $[\Phi_+(\eta_{\text{c}},\vec{x}),\Phi_-(\eta_{\text{c}},\vec{y})]$, and therefore do not contribute to fully connected scattering processes.\\
\indent Integrating by parts, using the equations of motion, and discarding all surface terms, the amplitude then becomes
\begin{align}
\langle \pm|\prod_i \mathcal{O}_\pm(\vec{x}_i)|\text{BD}\rangle&=\int\prod_i d^{d+1}x'_i\, \sqrt{-g}K_\pm (x_i',\vec{x}_i)\langle \pm |\text{T}\prod_i\Phi(x'_i)|\text{BD}\rangle_{\text{A}}.\label{simple}
\end{align}
where the subscript ``A" indicates that the external legs have been amputated. This provides a simple relation between truncated bulk correlation functions and matrix elements of the operators in the CFT. Note that the oscillatory surface terms at early times in (\ref{simple}) have been killed by a contour rotation (or an infrared regulator, from the point of view of the CFT). Indeed, $K_{\pm }$ is purely negative frequency with respect to the Bunch-Davies vacuum at the horizon, and only the negative frequency part of the $\Phi$ fields survives in the matrix element. Thus the entire integrand behaves as $e^{ik\eta}$ in the infinite past, and vanishes after taking $\eta\mapsto \eta+i\epsilon$. \\
\indent We have been referring to (\ref{amplitude}) as the S-matrix, but this is a slight misnomer -- a more accurate term would be the S-vector \cite{susskind}, since the initial state is fixed to be the Bunch-Davies vacuum. It would be interesting to understand how modifications of the initial state are reflected in the CFT.
\section{In-In Expectation Values}
\label{secinin}
So far, we have only considered transition amplitudes from the Bunch-Davies vacuum to some final state. In fact, these amplitudes contain the same information as the independent behaviors of Euclidean in-in expectation values near $\mathcal{I}^+$ (see \cite{weinberg} for a comprehensive review of the in-in formalism). To show this, let $F[\mathcal{O}_+]$ and $G[\mathcal{O}_-]$ be functions of $\mathcal{O}_{\pm}$. Inserting a complete set of states and using (\ref{generating}), we may express the expectation value of the product of $F$ and $G$ as
\begin{align}
\langle F[\mathcal{O}_+]G[\mathcal{O}_-]\rangle_{\text{BD}}&=\int [d\phi_+]\, \langle \text{BD}|\phi_+\rangle\langle \phi_+|F[\mathcal{O}_+]G[\mathcal{O}_-]|\text{BD}\rangle\notag\\
&=\int [d\phi_+]\, Z_{\text{CFT}}^*[\phi_+ ] F\left[\frac{\delta}{\delta \phi_+}\right]\left(G[2i\nu\phi_+]Z_{\text{CFT}}[\phi_+ ]\right),\label{inin}
\end{align}
where $Z_{\text{CFT}}[\phi_+]$ is the CFT partition function with the source $\phi_-$ turned off. By (\ref{postulate}), the expectation values (\ref{inin}) determine the behavior of expectation values of $\Phi$ near the boundary, so this behavior is in turn completely determined by the amplitudes (\ref{generating}). Note that insertions of $\mathcal{O}_-$ and $\mathcal{O}_+$ respectively encode the leading and subleading independent behaviors of expectation values at late times. \\
\indent The relation (\ref{inin}) does not treat $\mathcal{O}_+$ and $\mathcal{O}_-$ on an equal footing; expectation values of $\mathcal{O}_+$ are computed by functionally differentiating, while expectation values of $\mathcal{O}_-$ are evaluated by inserting a $\phi_+$ field. To remedy this, let us Legendre transform the CFT partition function,
\begin{align}
Z_{\text{CFT}}[\phi_+]=\int [d\phi_-]\, \exp\left(-2i\nu\int \frac{d^d\vec{k}}{(2\pi)^d}\phi_+(\vec{k})\phi_-(-\vec{k})\right)\widetilde{Z}_{\text{CFT}}[\phi_-].
\end{align}
Plugging this into (\ref{inin}), we then find the simple relation
\begin{align}
\langle F[\mathcal{O}_+]G[\mathcal{O}_-]\rangle_{\text{BD}}&=\int [d\phi_+\, d\phi_-]\, F\left[\frac{\delta}{\delta \phi_+}\right]G\left[-\frac{\delta}{\delta \phi_-}\right]\exp\left(-2i\nu\int \frac{d^d\vec{k}}{(2\pi)^d}\phi_+(\vec{k})\phi_-(-\vec{k})\right)\notag\\
&\hspace{25 mm}\times Z_{\text{CFT}}^*[\phi_+] \widetilde{Z}_{\text{CFT}}[\phi_-],\label{double}
\end{align}
which is more convenient for computations than (\ref{inin}). So we see that Bunch-Davies expectation values of $\mathcal{O}_\pm$ are computed by a system of two fluctuating fields $\phi_{\pm}$. A related approach to computing extrapolated Bunch-Davies expectation values is the doubled boundary theory of \cite{harlow2}.\\
\indent Note that the Legendre transform of the CFT partition function has a natural interpretation. In the AdS/CFT context, Legendre transforming the partition function is accomplished by switching to alternate boundary conditions on the bulk field \cite{klebanov}. Analytically continuing the resulting bulk correlation functions to de Sitter space gives the transition amplitudes
\begin{align}
\langle -|\text{T}\Phi(x_1)\cdots \Phi(x_n)|\text{BD}\rangle=\int_{\phi(\eta=-\infty)\sim e^{ik\eta}}^{\phi(\eta=0)\sim(-\eta)^{\Delta_-}}[d\phi]\, \phi(x_1)\cdots \phi(x_n)\exp(iS[\phi]).\label{transitionalt}
\end{align}
Thus the Legendre transform $\widetilde{Z}_{\text{CFT}}[\phi_-]$ is the generating functional for expectation values of $\mathcal{O}_-$ between $\langle -|$ and $\text{BD}\rangle$. \\
\indent In order to see how the above procedure works in practice, let us take a cubic bulk interaction,
\begin{align}
S_{\text{int}}=-\frac{\lambda}{3!}\int d^{d+1}x\, \sqrt{-g}\phi^3.
\end{align}
The terms in the wavefunction in (\ref{double}) that contribute to the tree-level three-point functions are then
\begin{align}
&Z_{\text{CFT}}[\phi_+]=\exp\left(\int \frac{d^d\vec{k}}{(2\pi)^d}\,\frac{i(ik)^{2\nu}\Gamma(1-\nu)}{4^{\nu}\Gamma(\nu)} |\phi_+(\vec{k})|^2\right.\notag\\
&\left.\hspace{40 mm}-\frac{i\lambda}{3!}\int d^{d+1}x' \sqrt{-g}\prod_{i=1}^{3}d^d\vec{x}_i\,K_+(x',\vec{x}_i)\phi_+(\vec{x}_i)\right),
\end{align}
and the Legendre transform $\widetilde{Z}_{\text{CFT}}$ is identical but with $\nu$, $K_+$, and $\phi_+$ replaced with $-\nu,$ $K_-$, and $\phi_-$. Performing the integrals in (\ref{double}), we find the CFT correlators
\begin{align}
\langle \mathcal{O}_\pm (\vec{k})\mathcal{O}_\pm (\vec{k}')\rangle_{\text{BD}}&=-\frac{\Gamma(1\mp \nu)^2 k^{\pm 2\nu}}{4^{\pm \nu}\pi}(2\pi)^d\delta^d(\vec{k}+\vec{k}'),\label{cftcor}\\
\langle \mathcal{O}_+ (\vec{k})\mathcal{O}_- (\vec{k}')\rangle_{\text{BD}}&=\nu(i-\cot(\nu\pi))(2\pi)^d\delta^d(\vec{k}+\vec{k}')\notag,\\
\left\langle \prod_{i=1}^{3}\mathcal{O}_{a_i}(\vec{k}_i)\right\rangle_\text{BD}&=-2i\lambda\nu^3\text{Im}\left(\int d\eta\, \sqrt{-g}\prod_{i}\frac{a_i\, K_{-a_i}(\eta,k_i)}{\text{Re}\langle -a_i|\mathcal{O}_{-a_i}(\vec{k}_i)\mathcal{O}_{-a_i}(-\vec{k}_i)|\text{BD}\rangle}\right)(2\pi)^d\delta^d\left(\sum_i \vec{k}_i\right)\notag,\end{align}
where each of the $a_i$'s can be equal to either $+$ or $-$. \\
\indent On the other hand, we may compute the two-point and three-point functions directly in the in-in formalism, 
\begin{align}
\langle \Phi(\eta,\vec{k})\Phi(\eta',\vec{k}')\rangle_{\text{BD}}&=\frac{\pi(\eta\eta')^{d/2}}{4}H^*_\nu(\eta k)H_\nu(\eta' k)(2\pi)^d\delta^d(\vec{k}+\vec{k}'),\\
\left\langle \prod_{i=1}^{3}\Phi(\eta,\vec{k}_i)\right\rangle_{\text{BD}}&=2\lambda \int {d\eta}\,\sqrt{-g}\, \text{Im }\left(\prod_{i}G_{\text{BD}}(k_i,\eta_{\text{c}},\eta)\right)(2\pi)^d\delta^d\left(\sum_i \vec{k}_i\right),\notag
\end{align}
where $G_{\text{BD}}$ is the Euclidean Wightman function. Pulling these results to the boundary then gives the same answers (\ref{cftcor}).
\section{Anomalies}
\label{secanomalies}
In this final section, we will investigate a phenomenon that was pointed out in \cite{maldacena,banks}: the term in the Hartle-Hawking wavefunction that is proportional to the Weyl anomaly drops out of the late-time probablity measure. \\
\indent To illustrate this point, let us consider three-dimensional gravity with a positive cosmological constant. We may compute the wavefunction by analytically continuing the partition function from AdS \cite{harlow}. Since the  central charge of the dual CFT to $AdS_3$ gravity is equal to $3\ell_{\text{AdS}}/(2G)$ \cite{brown}, where $\ell_{\text{AdS}}$ is the AdS radius, the AdS partition function takes the form\footnote{I would like to thank D. Harlow for explaining this argument.}  \cite{polyakov,polchinskibook}
\begin{align}
Z[\gamma]=\exp\left(-\frac{\ell_{\text{AdS}}}{32\pi G}\int d^2\vec{x}\, \sqrt{\gamma}\,d^2\vec{y}\, \sqrt{\gamma}\, R(\vec{x})\Box^{-1}R(\vec{y})\right)\label{psigravity},
\end{align}
where $R$ and $\Box$ are the Ricci scalar and Laplacian formed out of the boundary metric $\gamma$. Here we are assuming that $\gamma$ is topologically trivial, so that it is conformal to the flat metric $\delta_{ab}$. Performing the analytic continuation $\ell_{\text{AdS}}=i\ell_{\text{dS}}$ on (\ref{psigravity}), we find the Bunch-Davies wavefunction
\begin{align}
\Psi_{\text{BD}}[\gamma]=\exp\left(-\frac{i\ell_{\text{dS}}}{32\pi G}\int d^2\vec{x}\, \sqrt{\gamma}\,d^2\vec{y}\, \sqrt{\gamma}\, R(\vec{x})\Box^{-1}R(\vec{y})\right).\label{psibd}
\end{align}
We see from (\ref{psibd}) that the conformal anomaly cancels out when computing the late-time probability measure $|\Psi_{\text{BD}}[\gamma]|^2$. This implies that if matter fields are added to the theory in order to obtain a dynamical metric, then bulk expectation values are protected from the conformal anomaly. \\
\indent In fact, this observation extends to the matter anomaly as well. For instance, the conformal anomaly in AdS/CFT for a bulk scalar corresponding to an operator of dimension $\Delta_+=d/2+n$ for $n$ a positive integer is \cite{skenderis}
\begin{align}
A[\phi]\propto \ell_{\text{AdS}}^{d-1}z_{\text{c}}^{2n-d} \int\frac{d^d\vec{k}}{(2\pi)^d}k^{2n} |\phi(z_{\text{c}},\vec{k})|^2,
\end{align}
up to a real numerical constant. Under the analytic continuation $(z_{\text{c}},\ell_{\text{AdS}})\mapsto (-i\eta_{\text{c}},i\ell_{\text{dS}})$, this term picks up an overall imaginary factor of $i(-1)^{d-n+1}$,  implying that the anomaly again cancels out after squaring the wavefunction. \\
\indent Recall that the bulk interpretation of the conformal anomaly is that infrared divergences break the invariance of the bulk theory under the de Sitter isometry group $SO(d+1,1)$.  These divergences are consequences of the fact that de Sitter space has infinite volume, and should not be confused with the well-known IR divergences for a massless scalar in de Sitter space, which arise from the integration over zero modes in the propagator \cite{linde}. Thus, at least in the cases that we have considered here, late-time Euclidean expectation values of bulk fields are protected from the infinite-volume IR divergences in the wavefunction that break the de Sitter symmetries. This restoration of de Sitter invariance has been confirmed in a simple model of inflation \cite{freivogel}.\\
\indent It is natural to ask if expectation values of the conjugate momentum to $\phi$ are also protected from the conformal anomaly. To see that this is not the case, note that late-time expectation values of the conjugate momentum $\Pi$ can be computed by differentiating the wavefunction with respect to $\phi_0$ and integrating over $\phi_0$, 
\begin{align}
\langle F[\Pi(\eta_{\text{c}} ,\vec{x})]\rangle_{\text{BD}}&=\int [d\phi_0]\,|\Psi_{\text{BD}}[\phi_0,\eta_{\text{c}}]|^2F\left(\Psi_{\text{BD}}^{-1}[\phi_0,\eta_{\text{c}}]\frac{\delta \Psi_{\text{BD}}[\phi_0,\eta_{\text{c}}]}{i\delta \phi_0(\vec{x})}\right)\label{pi}.
\end{align}
This follows directly from Hamilton-Jacobi theory; the variation of the on-shell action with respect to the final boundary condition is equal to the conjugate momentum evaluated at the boundary. It is clear from (\ref{pi}) that the phase of the wavefunction does not drop out of expectation values of $\Pi$. Equivalently, the phase of the wavefunction contributes to Bunch-Davies expectation values of $\phi$ fields at different times, as noted in \cite{maldacena}. Naively this implies that a metaobserver \cite{witten} can compute the conformal anomaly by measuring late-time expectation values of $\Pi$. It would be interesting to understand if this effect is indeed physical, or if it can be eliminated via field redefinitions.
\section{Conclusion}
We have shown that computations in perturbative quantum field theory in de Sitter space can be reorganized as calculations in a hypothetical conformal field theory living at $\mathcal{I}^+$. One immediate application of our results is at a practical level: many of the difficult integrals involved in computing AdS/CFT correlators have been evaluated in the literature, and these can be carried over to Euclidean expectation values in de Sitter space by analytically continuing and using the formalism of Section \ref{secinin}. In addition, we expect that the formula (\ref{double}) is helpful for constraining possible shapes of the independent behaviors of late-time expectation values of massive fields, since the constraints imposed by conformal invariance are most easily analyzed from a wavefunction perspective \cite{maldacenapim}. \\ 
\indent Another avenue for future investigation would be to better understand the relation between our approach and that of \cite{strominger, strominger2, strominger3}. In those works, the bulk theory is dual to two different field theories on the boundary, depending on whether one imposes Dirichlet or Neumann boundary conditions on the bulk field at $\mathcal{I}^+$. Here we have chosen to impose future boundary conditions by fixing the final state rather than by fixing the asymptotic behavior of the bulk field, since this makes the connection to standard field theory in de Sitter space more transparent; however, it would be interesting to understand the latter perspective as well.\\
\indent On a broader note, any complete treatment of de Sitter holography must address the question of the finiteness of the entropy (see \cite{anninos} for a review). The bulk dual of our CFT Hilbert space is infinite-dimensional in perturbation theory, since one can insert any number of operators at arbitrary locations on the boundary. Recent investigations suggest that nonperturbative effects may reduce the number of states in the bulk \cite{strominger2}, but it is clear that much work still remains to be done.
 \\ \\ \textbf{Acknowledgements:} I am grateful to A. Jevicki for his guidance throughout the completion of this project, and for many valuable discussions. I would also like to thank D. Harlow and J. Polchinski for helpful discussions on related issues, and the anonymous referee for many insightful comments.
 
\end{document}